\documentstyle[12pt]{article}
\title{Hartree-Fock study of electronic ferroelectricity in the
Falicov-Kimball model with $f$-$f$ hopping}
\author{Pavol Farka\v sovsk\'y\\
Institute  of  Experimental  Physics,  Slovak   Academy   of
Sciences\\
Watsonova 47, 040 01 Ko\v {s}ice, Slovakia}
\date{}
\begin{document}
\baselineskip=24pt
\maketitle

\begin{abstract}

The Hartree-Fock (HF) approximation with the charge-density-wave (CDW) instability
is used to study the ground-state phase diagram of the spinless Falicov-Kimball
model (FKM) extended by $f$-$f$ hopping in two and three dimensions. It is 
shown that the HF solutions with the CDW instability reproduce perfectly the
two-dimensional intermediate coupling phase diagram of the FKM model with 
$f$-$f$ hopping calculated recently by constrained path Monte Carlo
(CPMC) method. Using this fact we have extended our HF study on cases
that have been not described by CPMC, and namely, (i) the case of small
values of $f$-electron hopping integrals, (ii) the case of weak Coulomb 
interactions and (iii) the three-dimensional case. We have found that
ferroelectricity remains robust with respect to the reducing strength of
coupling ($f$-electron hopping) as well as with respect to the increasing
dimension of the system.
\end{abstract}

\newpage

The Falicov-Kimball model (FKM) is a paradigmatic example of simple
model to study correlation effects in interacting fermion systems
on a lattice~\cite{Falicov}. The model was originally proposed to 
describe metal-insulator transitions and has since been investigated 
in connection with a variety of problems such as binary 
alloys~\cite{Freericks}, the formation of ionic crystals~\cite{Gruber}, 
and ordering in mixed-valence systems~\cite{Farkas}. 
In the last few years the FKM was extensively studied in connection with 
the exciting idea of electronic ferroelectricity~\cite{Cz,F1,Zl,B1,B2,Yin}. 
The motivation for these studies comes from the pioneering work of 
Portengen at al.~\cite{P1,P2} who studied the FKM with a k-dependent 
hybridization in the Hartree-Fock (HF) approximation and found 
that the Coulomb interaction $U$ between the itinerant $d$-electrons 
and the localized $f$-electrons gives rise a non-vanishing excitonic
$\langle f^{+}d\rangle$-expectation value even in the limit of vanishing
hybridization $V \rightarrow 0$. As an applied (optical) electrical field
provides for excitations between d- and f-states and thus for a polarization
expectation value $P_{fd}=\langle f_i^{+}d_i\rangle$, the finding of a spontaneous
$P_{fd}$ (without hybridization or electric field) has been interpreted as
evidence for electronic ferroelectricity. This result stimulated further
theoretical studies of the model. 
Analytical calculations within well controlled approximation (for
$U$ small) performed by Czycholl~\cite{Cz} in infinite dimensions did
not confirm the existence of electronic ferroelectricity. In contrast to 
results obtained by Portengen et al.~\cite{P1,P2} he found that the FKM
in the symmetric case ($n_f=n_d=0.5$) does not allow for a 
ferroelectric ground state with a spontaneous polarization,
i.e., there is no nonvanishing $\langle f^{+}d\rangle$-expectation value in the
limit of vanishing hybridization.  The same conclusion has been also 
obtained independently by extrapolation of small-cluster 
exact-diagonalization and density matrix renormalization group (DMRG) 
calculations in the one dimension for both 
intermediate and strong interactions~\cite{F1}. In these regions 
the finite-size effects are negligible and thus the results can be 
satisfactory extrapolated to the thermodynamic limit. 

Hybridization between the itinerant $d$ and localized $f$ states, however,
is not the only way to develop $d$-$f$ coherence. Recent theoretical works
of Batista et al.~\cite{B1,B2} showed that the ground state with 
a spontaneous electric polarization can also be induced by $f$-$f$ hopping
for dimensions $D>1$.
In the strong coupling limit this result has been proven by mapping the
extended FKM into the $xxz$ spin 1/2 model with a magnetic field along 
the $z$-direction, while in the intermediate coupling regime the
ferroelectric state has been identified numerically by constrained path 
Monte Carlo (CPMC) technique. 
On the base of these results the authors postulated 
the following conditions that favor the formation of the electronically 
driven ferroelectric state: (a) The system must be in a mixed-valence regime 
and the two bands involved must have different parity. (b) It is best, 
though not necessary, if both bands have similar bandwidths. (c) A local 
Coulomb repulsion ($U$) between the different orbitals is required. 

In the present paper we study the extended FKM (the spinless FKM with
$f$-$f$ hopping) in the HF approximation with the charge-density-wave (CDW) 
instability. For reasons mentioned above we restrict our studies
on dimensions $D>1$. First we show that the HF solutions
with the CDW instability reproduce perfectly the ground-state phase 
diagram obtained by the CPMC method ($D=2$) for 
intermediate Coulomb interactions~\cite{B2}. This "calibration" allows us to 
extend calculations to the case of small values of the $f$-electron hopping
integral $|t_f| < 0.1$, that has been omitted in the CPMC phase diagram 
for numerical problems. Just in this region 
we have found a new phase that corresponds to the inhomogeneous solution    
for $\langle f^+d\rangle$-expectation value. This result completes the ground-state 
phase diagram of the two-dimensional FKM extended by $f$-$f$ hopping for 
intermediate couplings. The same calculations we have performed also in 
the weak coupling limit (for $D=2$) as well as in three dimensions.  
We have found that the ferroelectricity remains robust with respect to the
reducing strength of the coupling as well as with respect to the increasing
dimension of the system. 

\section{The model}

The extended FKM for the spinless fermions on a $D$-dimensional hypercubic
lattice is
\begin{equation}
H=-t_d\sum_{<ij>}d^+_id_j-t_f\sum_{<ij>}f^+_if_j+U\sum_if^+_if_id^+_id_i
+E_f\sum_if^+_if_i,
\end{equation}
where $f^+_i$ $(f_i)$ and $d^+_i$ $(d_i)$
is the creation (annihilation) operator  of heavy ($f$)  and light ($d$) 
electron at lattice site $i$.

The first two terms of (1) are the kinetic energies corresponding to
quantum-mechanical hopping of $d$ and $f$ electrons 
between the nearest neighbor sites $i$ and $j$
with hopping probabilities $t_d$ and $t_f$, respectively.
The third term represents the on-site
Coulomb interaction between the $d$ electrons with density
$n_d=\frac{1}{L}\sum_id^+_id_i$ and the $f$ electrons with density 
$n_f=\frac{1}{L}\sum_if^+_if_i$, where $L$ is the number of lattice sites. 
Usually, the hopping integral of the $d$ electrons is taken to be the unit 
of energy $(t_d=1)$ and the $f$-electron hopping integral is considered in the 
limit $|t_f| < 1$. This is a reason why the $d$ electrons are called
light and the $f$ electrons heavy.

In our HF study of the extended FKM we go beyond the usual HF
approach~\cite{Leder} in which
only homogeneous solutions are postulated. In accordance with~\cite{Brydon} 
we consider here also inhomogeneous solutions modeled by  a periodic 
modulation of the order parameters:
\begin{eqnarray}
\langle n_i^f\rangle=n^f+\delta_f\cos({\bf Q}\cdot {\bf r}_i)\ ,\\
\langle n_i^d\rangle=n^d+\delta_d\cos({\bf Q}\cdot {\bf r}_i)\ ,\\
\langle f_i^+d_i\rangle=\Delta+\Delta_P\cos({\bf Q}\cdot {\bf r}_i)\ .
\end{eqnarray}
where $\delta_{d}$ and $\delta_{f}$ is the order parameter of the CDW state 
for the $d$- and $f$-electrons and $\Delta$ is the excitonic average.
The nesting vector ${\bf{Q}}=(\pi,\pi)$ for $D=2$ and 
${\bf{Q}}=(\pi,\pi,\pi)$ for $D=3$.

Using expressions for $\langle n_i^f\rangle$,$\langle n_i^d\rangle$ 
and $\langle f_i^+d_i\rangle$ the HF Hamiltonian of the extended
FKM can be written as
\begin{eqnarray}
{\cal H}=&-&t_d\sum_{\langle i,j\rangle} d^+_id_j - t_f\sum_{\langle i,j\rangle}f^+_if_j
+ E_f\sum_i n_i^f
      + U\sum_i (n^f+\delta_f\cos({\bf Q}\cdot {\bf r}_i))n_i^d
\nonumber\\
      &+& U\sum_i (n^d+\delta_d\cos({\bf Q}\cdot {\bf r}_i))n_i^f
      -U\sum_{i} ( \Delta+\Delta_P\cos({\bf Q}\cdot {\bf r}_i))
      d_i^+f_i+h.c. 
\end{eqnarray}
Following the work of Brydon et al.~\cite{Brydon} the effective HF Hamiltonian
is diagonalized by canonical transformation 
\begin{eqnarray}
\begin{array}{ccc}
\gamma_k^m=u_k^md_k + v_k^md_{k+{\bf Q}} + a_k^mf_k + b_k^mf_{k+{\bf Q}}\ , &   &
m=1,2,3,4\ ,
\end{array}
\end{eqnarray}
where $a_k^m,b_k^m,u_k^m,v_k^m$ are solutions of the associated Bogoliubov-de Gennes (BdG)
eigenequations:

\begin{equation}
H_k\Psi_k^m=E_k^m\Psi_k^m\ , 
\end{equation}
with
\begin{eqnarray}
H_k=\left( \begin{array}{cccc}
    \epsilon_k^d+Un^f &  U\delta_f &   -U\Delta &     -U\Delta_P \\
     U\delta_f  & \epsilon_{k+Q}^d+Un^f & -U\Delta_P &    -U\Delta\\
     -U\Delta   &  -U\Delta_P   &  \epsilon_{k}^f+Un^d+E_f & U\delta_d \\
     -U\Delta_P &  -U\Delta    &  U\delta_d  & \epsilon_{k+Q}^f+Un^d+E_f\\
            \end{array}          
\right)
\end{eqnarray}
and
\begin{eqnarray}
\Psi_k^m=\left(\begin{array}{c}
               u_k^m\\
               v_k^m\\
               a_k^m\\
               b_k^m
                \end{array}
\right) 
\end{eqnarray}

The corresponding energy dispersions $\epsilon_k^d$ and $\epsilon_k^f$ 
can be obtained directly by the Fourier transform of the $d$- and
$f$-electron hopping amplitudes and for the case of hypercubic lattice
they are given by ($\alpha=d,f$):

\begin{eqnarray}
\epsilon_k^{\alpha}&=&-2t_{\alpha}(\cos(k_x)+\cos(k_y)), \mbox{ for D=2},\\
\epsilon_k^{\alpha}&=&-2t_{\alpha}(\cos(k_x)+\cos(k_y)+\cos(k_z)), \mbox{
for D=3}.
\end{eqnarray}

The HF parameters $n_d,\delta_d,n_f,\delta_f,\Delta,\Delta_P$ can be written
directly in terms of the Bogoliubov-de Gennes eigenvectors:
\begin{equation}
n^d=\frac{1}{N}\sum_k{}'\sum_m \{ u_k^mu_k^m + v_k^mv_k^m\}f(E_k^m)\ .
\end{equation}
\begin{equation}
\delta_d=\frac{1}{N}\sum_k{}'\sum_m \{ v_k^mu_k^m + u_k^mv_k^m\}f(E_k^m)\ .
\end{equation}
\begin{equation}
n^f=\frac{1}{N}\sum_k{}'\sum_m \{ a_k^ma_k^m + b_k^mb_k^m\}f(E_k^m)\ .
\end{equation}
\begin{equation}
\delta_f=\frac{1}{N}\sum_k{}'\sum_m \{ b_k^ma_k^m + a_k^mb_k^m\}f(E_k^m)\ .
\end{equation}
\begin{equation}
\Delta=\frac{1}{N}\sum_k{}'\sum_m \{ a_k^mu_k^m + b_k^mv_k^m\}f(E_k^m)\ .
\end{equation}
\begin{equation}
\Delta_P=\frac{1}{N}\sum_k{}'\sum_m \{ b_k^mu_k^m + a_k^mv_k^m\}f(E_k^m)\ .
\end{equation}
where the  prime denotes summation over half the Brillouin zone and 
$f(E)=1/\{1+\exp[\beta(E-\mu)]\}$ is the Fermi distribution function.

The same approach has been used recently by Brydon et al.~\cite{Brydon} 
to study the interplay between excitonic effects and the CDW instability in the FKM with  on-site as well as 
non-local hybridization. Here we use the zero temperature variant of this
procedure to describe ground-state phase diagram of the spinless
FKM with $f$-$f$ hopping.

\section{Results and Discussion}
To determine the ground-state phase diagram of the extended FKM in the 
$E_f$-$t_f$ plane (corresponding to selected $U$) the HF equations are solved 
self-consistently for each pair of $(E_f,t_f)$ values.
We use an exact diagonalization method to solve the Bogoliubov-de Gennes  
equation. We start with an initial set of order parameters. 
By solving Eq.~(7), the new
order parameters are computed via Eqs.~(12) to
(17) and are substituted back into
Eq.~(7). The iteration is repeated until a desired
accuracy is achieved.

First we have examined the two-dimensional extended FKM model in the 
intermediate coupling regime and $t_f$ negative. For this case there 
exists the 
comprehensive phase diagram of the model obtained by a CPMC 
technique~\cite{B2}  for $f$-electron hopping integrals $|t_f| \geq 0.1$.
According these Monte-Carlo studies the phase diagram of the extended FKM
consists of only three main phases, and namely, (i) the integer-valent 
state ($n_f=0,1, n_d=1,0$), (ii) the mixed-valent CDW state ($n_f=n_d=0.5$), 
and (iii) the mixed-valent ferroelectric state that is stable for remaining
values of $n_f$ ($n_d$).     

In Fig.~1 we have displayed typical examples of our HF solutions 
obtained for $n_d,\delta_d,n_f,\delta_f,\Delta,\Delta_P$ in the intermediate 
coupling regime $U=2$. 
It is seen that the extended FKM in the HF approximation with the
CDW instability exhibits non-vanishing excitonic  $\langle
f^+d\rangle$-expectation value for all $f$-electron densities except the
case when $n_f=0, 1/2$ and 1. Thus in accordance with the Quantum
Monte-Carlo studies~\cite{B2} we have found that the ferroelectric
ground state with the spontaneous polarization is stabilized when the system
is in the mixed valence regime and the sign of the $f$-electron hopping
integral is opposite to the sign of the $d$-electron one. The fact that
HF solutions can describe the existence of ferroelectric
ground-state with spontaneous polarization is not surprising, since this
state has been found already in the homogeneous HF solution of the
conventional FKM ($t_f=0$) in the limit of vanishing hybridization 
$V\rightarrow 0$~\cite{P1,P2}, even for all
$f$-electron concentrations (for all values of $E_f$ from the $d$-electron
band). What is however surprising is that the HF solutions with the CDW
instability reproduce perfectly the ground-state phase diagram obtained by
CPMC method for all examined values of $f$-electron hopping ($|t_f|\geq
0.1$). This is clearly demonstrated in Fig.~2, where both phase diagrams are
compared.

The fact that the HF approximation with the CDW instability can describe
qualitatively as well as quantitatively ground-state properties of the FKM
with $f$-electron hopping motivated us to extend our HF study on cases that 
have been not described by Quantum Monte-Carlo simulations. At first 
this is the case of
small $f$-electron hopping integrals ($|t_f|<0.1$) that has been not
considered in the original work of Batista et al.~\cite{B2} because 
numerical difficulties which appear in the Quantum Monte-Carlo simulations
for small $t_f$ (the limitations in the numerical accuracy). The second
interesting case that we would like to study here within the HF theory is
the three-dimensional case for which the numerical results are very rare due
to numerical limitations on the size of clusters. 

Let us first discuss our two-dimensional results obtained in the limit of
small values of $f$-electron hopping integrals. In Fig.~3 we present
results of detailed HF analysis  performed in this limit for 
$\Delta,\Delta_P$ and $\delta_d$. It is seen that the non-vanishing 
excitonic $\langle f^+d\rangle$
expectation value persists also for small values of $|t_f|$ but now the
inhomogeneous solution $\Delta_P\neq 0$ (with AB-sublattice oscillations in
the excitonic and charge order parameters) is stabilized
against the homogeneous one ($\Delta_P=0$). The effect is especially strong
when we approach the $t_f=0$ limit. This is clearly demonstrated in Fig.~4,
where the complete intermediate-coupling phase diagram of the FKM with
$f$-$f$-hopping is displayed. Five different phases depicted in Fig.~4 as 
$\alpha$ (the full $f$ band), $\beta, \beta'$ (the excitonic phases),
$\gamma$ (the CDW phase) and $\epsilon$ (the full $d$ band) correspond
to following HF solutions: 
\begin{eqnarray}
 \begin{array}{cccccll}
  \alpha \mbox{ phase:} & \Delta=0, & \Delta_P=0,   
  & \delta_f=0, & \delta_d=0,  & n_f=1
  \nonumber\\
  \beta \mbox{ phase:} & \Delta>0, &  \Delta_P=0,   
  & \delta_f=0, & \delta_d=0,  & 0<n_f<n_f^c & \mbox{for $E_f>0$} 
  \nonumber\\
    & &  
  & &  & 1-n_f^c<n_f<1 & \mbox{for $E_f<0$} 
  \nonumber\\
  \beta' \mbox{ phase:} & \Delta > 0, & \Delta_P< 0,   
  & \delta_f<0, & \delta_d>0, & n_f^c<n_f<1/2 & \mbox{for $E_f>0$}  
  \nonumber\\
    & &  
  & &  & 1/2<n_f<1-n_f^c & \mbox{for $E_f<0$} 
  \nonumber\\
  \gamma \mbox{ phase:} & \Delta=0,  &\Delta_P=0,   
  & \delta_f<0, & \delta_d>0,  & n_f=1/2
  \nonumber\\
  \varepsilon \mbox{ phase:} & \Delta=0, & \Delta_P=0,   
  & \delta_f=0, & \delta_d=0,  & n_f=0
  \nonumber\\
 \end{array}
\end{eqnarray}
The stability of different HF solutions was also checked numerically by
calculating the total energy and it was found that all phases presented in
the ground-state phase diagram represent the most stable HF solutions. 
To determine the type of transitions between different phases we have
performed an exhaustive numerical study of the $E_f$ dependence of the HF
order parameters (the typical examples are shown in Fig.~1 and Fig.~3). 
At a first glance it seems that there are both first-order ($t_f$ large) and 
second-order ($t_f$ small) phase transitions in the extended FKM with $f$-$f$
hopping. However, a more detailed analysis of numerical data (with much 
higher resolution than used in Fig.~1 and Fig.~3) showed that the $\beta'$ 
phase persists also for large $t_f$, although its stability region is now 
considerably reduced (see insets in Fig.~1). Thus there is no difference 
between the case of small and large values of $t_f$. In both cases the HF
order parameters change continuously indicating that the phase transitions
between different phases presented in the ($E_f$-$t_f$) ground-state phase 
diagram are of the second order.

The same calculations we have performed also in the weak coupling limit
($U \leq 1$). We have found that the phase diagrams obtained in the weak and
intermediate coupling regime have qualitatively the same form and only one
difference between them is that the ferroelectric domain ($\beta$) is 
stabilized against remaining phases with decreasing Coulomb 
interaction (see Fig.~4). Of course, this fact does not imply automatically 
that the excitonic $\langle f^+d\rangle$ expectation value persists also 
for vanishing $U$ and that the Coulomb interaction $U$ is not necessary 
for a stabilization of the ferroelectric state, what should be in 
a contradiction with conclusions  based on the CPMC simulations. 
Indeed, calculations that we have performed for different values of $t_f$ 
at the selected $f$-electron density $n_f=1/4$ showed (see Fig.~5) that the excitonic 
$\langle f^+d\rangle $ expectation  value is zero for $U=0$,
rapidly increases with increasing $U$ and tends to the saturated state for
$U$ sufficiently large. This confirms independently the third postulate of
Batista et al.~\cite{B2} and namely, that the local Coulomb interaction 
between the different orbitals is required in order to stabilize the 
ferroelectric state with the spontaneous polarization.

Before discussing the case of positive $t_f$ let us show explicitly the HF
solution for the limit of the conventional FKM ($t_f=0$). For this case we 
have found that $\Delta=\Delta_P=0$ in the $\alpha, \gamma$ and $\varepsilon$ 
phase, while $\Delta=-\Delta_P$
in the $\beta'$ phase. The last solution implies that the extitonic $\langle
f^+_id_i\rangle$-expectation value is equal to $2\Delta$ on the $A$
sublattice of the hypercubic lattice, while $\langle f^+_id_i\rangle =0$ on
the $B$ sublattice. For the symmetric case $E_f=0$ our solutions are fully
consistent with the Czycholl's ones obtained in the limit of infinite
dimensions~\cite{Cz}. On the other hand both these inhomogeneous solutions
fully differ from the homogeneous one~\cite{P1,P2} that predicts a non-zero
excitonic $\langle f^+_id_i\rangle$-expectation value for all $E_f$ from the
mixed valence regime with maximum of $\langle f^+_id_i\rangle$ at $E_f=0$.

Similar calculations as for $t_f<0$ we have performed also for $t_f>0$. We
have found that the ground-state phase diagram for $t_f>0$  has exactly the
same form as for $t_f<0$, however five different phases $\alpha, \beta, 
\beta', \gamma$ and $\varepsilon$
are now characterized by:
\begin{eqnarray}
 \begin{array}{cccccll}
  \alpha \mbox{ phase:} & \Delta=0, & \Delta_P=0,   
  & \delta_f=0, & \delta_d=0, & n_f=1
  \nonumber\\
  \beta \mbox{ phase:} & \Delta = 0, & \Delta_P < 0,   
  & \delta_f=0,  & \delta_d=0, & 0<n_f<n_f^c  & \mbox{for $E_f>0$}
  \nonumber\\
   & & 
  &  & & 1-n_f^c<n_f<1  & \mbox{for $E_f<0$}
  \nonumber\\
  \beta' \mbox{ phase:} & \Delta > 0, & \Delta_P< 0,   
  & \delta_f<0, & \delta_d>0, & n_f^c<n_f<1/2 &  \mbox{for $E_f>0$}
  \nonumber\\
   & & 
  &  & & 1/2<n_f<1-n_f^c  & \mbox{for $E_f<0$}
  \nonumber\\
  \gamma \mbox{ phase:} & \Delta=0,  &\Delta_P=0,   
  & \delta_f<0, & \delta_d>0, & n_f=1/2
  \nonumber\\
  \varepsilon \mbox{ phase:} & \Delta=0, & \Delta_P=0,   
  & \delta_f=0, & \delta_d=0 & n_f=0
  \nonumber\\
 \end{array}
\end{eqnarray}
Thus the main difference between the phase diagrams obtained for 
negative and positive $t_f$ is that the ferroelectric domain $\beta$ at $t_f<0$ 
is replaced by the antiferroelectric one at $t_f>0$. These two large domains
are separated by a relatively narrow $\beta'$ domain within which the
sublattice excitonic averages ($P^{A}_{fd},P^{B}_{fd}$) change 
continuously (see Fig.~6) from the ferroelectric case 
($P^{A}_{fd}=P^{B}_{fd}$) to the antiferroelectric case 
($P^{A}_{fd}=-P^{B}_{fd}$). 

Qualitatively the same picture we have observed also for the
three-dimensional case. This is illustrated  in Fig.~7, where the
ground-state phase diagrams of the extended FKM are plotted for two different
values of Coulomb interaction ($U=2$ and $U=4$). These results indicate that
ferroelectricity remains robust with respect to the increasing dimension of
the system, what should be important for an application of HF
solutions on a description of real three dimensional systems.

In conclusion, we have calculated the ground-state phase diagram 
of the spinless FKM with $f$-$f$ hopping in the HF approximation 
with the CDW instability. We have found that the HF solutions with the CDW 
instability reproduce perfectly the two-dimensional intermediate coupling
phase diagram of the extended FKM calculated by CPMC method. Using this 
fact we have extended our HF study on cases that have been not described by 
CPMC and namely, the case of small values of $f$-electron hopping integrals, 
the case of weak Coulomb interactions and the three-dimensional case. 
We have found that the ferroelectric ground state with the spontaneous 
polarization remains stable in all examined cases.

\vspace{0.5cm}
This work was supported by the Slovak Grant Agency for Science
under grant No. 2/7057/27 and the Slovak APVV Grant Agency under Grant
LPP-0047-06. I would also like to acknowledge H. \v Cen\v carikov\'a for a
technical help during the preparation of manuscript.
\\
{\it Note added.-} After submitting this work we came to know about the work
of Schneider and Czycholl~\cite{Schneider} who studied the extended FKM in
the limit of infinite dimensions and obtained results similar to ours.

\newpage

\centerline{\bf Figure Caption}

\vspace{0.5cm}
Fig.~1. Dependence of the HF parameters $n_f,\delta_f,n_d,\delta_d,\Delta$
and $\Delta_P$ on the $f$-level energy $E_f$ calculated (with step $\Delta
E_f=0.005$) for three different values of $t_f$ ($t_f=-0.2, -0.5,-0.8$) 
and $U=2$. Insets show the $t_f=-0.5$ case at much higher resolution (the
numerical data have been obtained with step $\Delta E_f=0.00005$). The case
of $t_f=-0.8$ is analogous to $t_f=-0.5$.

\vspace{0.5cm}
Fig.~2. The HF ($\bullet$) and CPMC~\cite{B2} ($\Box$) phase diagram of the two
dimensional FKM with $f$-$f$ hopping obtained for $U=2$.

\vspace{0.5cm}
Fig.~3. Dependence of the HF parameters $\Delta$, $\Delta_P$ and $\delta_d$ 
on the $f$-level energy $E_f$ calculated for different values of $t_f$ 
($t_f=0, -0.01, -0.02, -0.05$) and $U=2$.

\vspace{0.5cm}
Fig.~4. The complete HF phase diagram of the two-dimensional extended FKM in
the intermediate ($U=2$) and weak coupling ($U=1$) regime. 

\vspace{0.5cm}
Fig.~5. Dependence of the HF parameter $\Delta$ on the Coulomb interaction
$U$ calculated for different values of $t_f$ and $n_f=1/4$.

\vspace{0.5cm}
Fig.~6. Dependence of the excitonic expectation value 
$P_{fd}=\langle f^{+}_id_i\rangle$ on $t_f$ calculated 
for $E_f=0.7$ and $U=2$.

\vspace{0.5cm}
Fig.~7. The complete HF phase diagram of the three-dimensional extended FKM
calculated for $U=2$ and $U=4$.

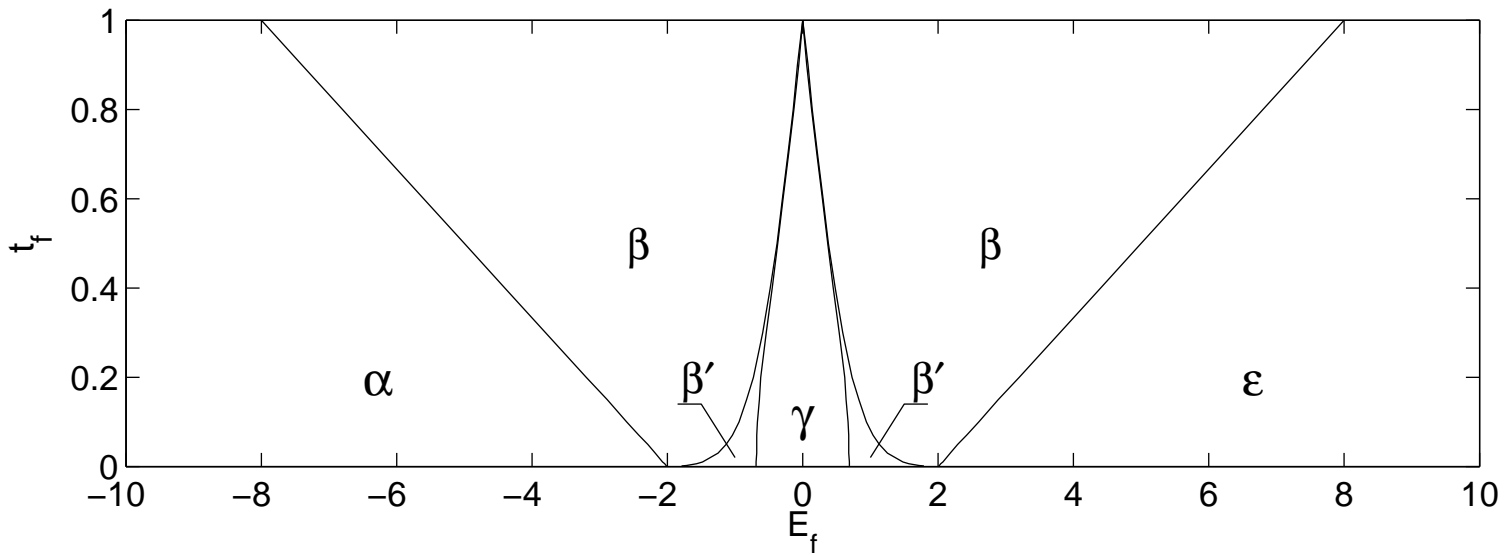
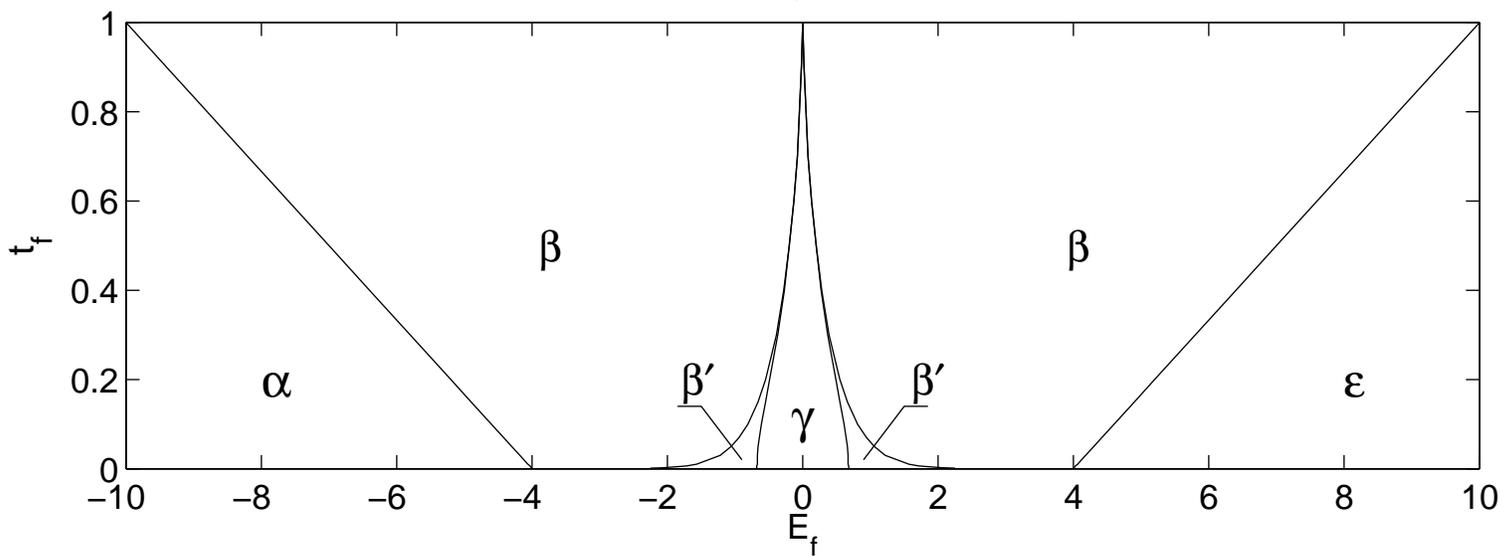
\end{document}